\documentclass[11pt]{article}
\input{amssym}

\newsavebox{\PSLASH}
\sbox{\PSLASH}{$p$\hspace{-1.8mm}/}

\begin{document}
\title{ SLE with Jumps and Conformal Null Vectors }
\author{S. Moghimi-Araghi\footnote{e-mail: samanimi@sharif.edu} , M. A. Rajabpour\footnote{e-mail: rajabpour@mehr.sharif.edu}
, S. Rouhani\footnote{e-mail: rouhani@ipm.ir} \\ \\
Department of Physics, Sharif University of Technology,\\ Tehran,
P.O.Box: 11365-9161, Iran} \date{} \maketitle

\begin{abstract}
Ordinary SLE$_{k}$ is defined using a Wiener noise and is related
to CFT's which have null vector at level two of conformal tower.
In this paper we introduce stochastic variables which are made up
of jumps and extend the ordinary  SLE to have such stochastic
variables.  The extended SLE can be related to CFT's which have
null vectors in higher levels of Virasoro module.
  \vspace{5mm}%
\end{abstract}

\section{Introduction}
Conformal field theories(CFT's) \cite{Polyakov} are powerful tools
to analyze critical behavior of two dimensional systems. In
addition, CFT has been remarkably successful in calculating the
geometrical properties of some geometrical critical phenomena such
as percolation, Brownian motion and so on. In recent years,
another approach to study geometrical models has become popular,
Schramm (Stochastic) Loewner evolution (SLE)\cite{Schramm}. SLE is
a one-parameter family of stochastic equations which describes the
growth of random curves. The parameter is usually named to be
$\kappa$ and hence these evolution are sometimes referred to
SLE$_\kappa$.

Some results, which were derived using CFT methods before, can be
recalculated by means of SLE. Among them is Cardy's
formula\cite{cardy}, which can be re-derived in a more general
form. As the two methods are able to calculate the same problems,
one concludes that there should be a close relation between CFT
and SLE. In \cite{bau2} and later in \cite{Friedrich} the relation
between CFT and SLE was proposed. The method used by \cite{bau2}
can be simply stated as follows: using Ito's formulas for
Loewner's equation and then extending the Witt algebra to Virasoro
algebra, an equivalent stochastic equation in Virasoro group is
found, and if the CFT model has null vector at level two, one is
able to find the martingales of SLE. In this case the area taken
into account is the upper half plane, but in \cite{rasmussen} it
was explained that if we work on some other domains, some special
null vectors in higher levels could be produced.

To find other null vectors, we introduce new stochastic variables,
which are essentially made up of jumps, and extend the drift term
of Schramm-Loewner equation to have these new variables. Doing
this, we are able to produce all of the operators of the Witt
algebra, so we can construct arbitrary null vectors.

In the following section we define SLE and then state its relation
to CFT briefly, more details can be found in
\cite{bau2,bau3,bau4,bau5}. In third section we introduce some
stochastic variables which can be replaced by the Wiener noise
present in ordinary SLE, and discuss their properties. The last
section is devoted to differential equations corresponding to null
vectors at arbitrary level.

\section{Chordal Stochastic Loewner Evolutions and CFT}
SLE, is an stochastic equation which reveals the evolution of
mappings from upper half plane, $H$ to itself. Therefore, this
equation could be regarded as the equation which gives the
evolution of the boundary. Let the mapping at time $t$ be
$g_t(z)$, then SLE evolution is written in the following form
\begin{equation}\label{SLEMain}
\partial_{t}g_{t}(z)=\frac{2}{g_{t}(z)-\xi_{t}},
\end{equation}
where $\xi_t$ is an stochastic variable. In general, $\xi_{t}$ can
be any function, but in standard SLE, it is taken to be Brownian
motion, that is $\langle\xi_{t}\xi_{s}\rangle=k \min(t,s)$, and
hence a one parameter family of such evolutions is defined, which
usually are referred to SLE$_k$.

At any time, $t$, $g_{t}(z)$ has a domain of analyticity which is
named $H_{t}$. This region is mapped to $H$ by the conformal
mapping $g_t(z)$. The complement of $H_{t}$ in $H$ is called the
hull in which the map is not well defined. As the times passes,
the hull grows, so for a fixed $z$, the mappings are well defined
up to the time $\tau_{z}\leq\infty$ with the property
$g_{\tau_z}(z)=\sqrt{k}B_{\tau_{z}}$. At this point, the
denominator of right hand side of  (\ref{SLEMain}) vanishes and
the mapping becomes meaningless. The trace of SLE is defined to be
the path of such singularities, that is,  $\gamma(t)= \lim_{z
\rightarrow 0} g_{t}^{-1}(z+\xi_{t})$. SLE has been studied a lot
in recent years and has found many different applications, for
reviews see for example \cite{Rhode,Lawler,kager,05Cardy}.

To establish the connection between SLE and CFT, we first define
the new series of maps $f_{t}(z)\equiv g_{t}(z)-\xi_{t}$ and write
SLE in the Langvien form which is more convenient.
\begin{equation}\label{SLEMain2}
df_{t}=\frac{2 dt}{f_{t}}-d\xi_{t}.
\end{equation}
This form of SLE helps us to write Ito's formula :
\begin{equation}\label{SLEMain3}
d\gamma_{f_{t}}.F=(\gamma_{f_{t}}.\acute{F})\left(\frac{2
dt}{f_{t}}-d
\xi_{t}\right)+\frac{f}{2}(\gamma_{f_{t}}.F^{\prime\prime}),
\end{equation}
where $\gamma_{f_{t}}.F\equiv F\circ f$.$\gamma_{f_{t}}$ belongs
to the group $N_{-}$ of germs of holomorphic functions at $\infty$
of the form $z+\sum_{m\leq-1}f_{m}z^{m+1}$. We can write the
equation(\ref{SLEMain3}),in the $N_{-}$ group space:
\begin{equation}\label{SLEMain4}
\gamma_{f_{t}}^{-1}d\gamma_{f_{t}}=2
dt\left(\frac{2}{z}\partial_{z}+\frac{k}{2}\partial_{z}^{2}\right)-d\xi_{t}\partial_{z}
\end{equation}
We observe that some differential operators in the form
$\ell_n=-z^{n+1}\partial_z$ have emerged. These operators form an
algebra called Witt algebra
\begin{equation}
[l_{n},l_{m}]=(m-n)l_{n+m}
\end{equation}
These operators are the generators of the conformal mapping in the
complex plane. On the quantum level, the corresponding algebra is
the well known Virasoro algebra with corresponding generators
$L_n$'s. So, the equation (\ref{SLEMain4}) in the Hilbert space
turns out to be
\begin{equation}\label{itovir}
G_{f_{t}}^{-1}dG_{f_{t}}=dt\left(-2L_{-2}+\frac{k}{2}L_{-1}^{2}\right)+d\xi_{t}L_{-1}.
\end{equation}
This equation defines stochastic trajectories on manifold of the
group produced by elements of Virasoro algebra which have some
martingales in verma module. To find these martingales we can make
both side of equation (\ref{itovir}) operate on the highest weight
vector of a representation of Virasoro algebra, $|\omega\rangle$,
with conformal weight $h=\frac{6-k}{2k}$ and central charge
$c=\frac{(6-k)3k-8)}{2k}$. As the vector
$-2L_{-2}+\frac{k}{2}L_{-1}^{2}|\omega\rangle$ is a null vector
and hence is orthogonal to all vectors in Verma module, one finds
\begin{equation}\label{SLEMain5}
E[G_{f_{t}}|\omega\rangle]=G_{f_{s}} |\omega\rangle,
\end{equation}
where the time averaging is for all times less than $s$. This
means that correlation functions of the conformal field theory in
H$_{t}$ are time independent and equal to their value at $t=0$.
Let's see what the state $G_{t}|\omega\rangle$ means. Suppose
$|\omega\rangle$ be a boundary changing operator in H, then one
can show that the equivalent operator in H$_{t}$ is just
$G_{t}|\omega\rangle$. In fact $G_{t}|\omega\rangle$ is a
generating function for all conserved quantities in chordal SLE.

We can repeat this calculation for SLE in $\pi/n$ space,
${\frac{H}{n}}$, with the modified Loewner's
equation\cite{rasmussen}:
\begin{equation}\label{Rasmuss}
\partial_{t}g_{t}(z)=\frac{2}{g_{t}(z)^{n-1}\left(g_{t}(z)^{n}-\xi_{t}\right)}.
\end{equation}
$g_{t}(z)$ maps the hull of $g_t(z)$, ${\frac{H_{t}}{n}}$, to the
whole ${\frac{H}{n}}$. To connect this modified equation to CFT,
one defines $f^{n}_{t}=g_{t}(z)^{n}-\xi_{t}$, and with similar
steps to derive martingales in Virasoro group for some $n$'s.

\section{Stochastic Jump Variables}
In the previous section, we introduced a class of stochastic
equation where the stochastic part came from a Wiener noise. In
this section, we'll introduce a new stochastic variable, which is
basically produced by a series of jumps, and show that in a
special case, this new noise will be identical with Brownian
motion. This helps us extend the stochastic Loewner equations
using the new stochastic variables.

Consider the  following stochastic equation
\begin{equation}\label{Jdef}
dx=J\;dN.
\end{equation}
The variable $dN$ is always zero except at some points, say at
$t=t_i$, where it takes the value 1. One can state this equation
in another form which may be more understandable
\begin{equation}\label{Xdefdel}
\frac{dx}{dt}=\sum_i J_i \delta(t-t_i).
\end{equation}
Now it is clear that the variable $x$, is constant. It has only
some jumps at times equal with $t_i$'s with magnitude $J_i$. We
have supposed that the magnitude of jumps can be different, in
general we can consider that jumps have the distribution function
$\rho(J)$. Also we will assume that the distribution of jumps in
time is a Poison distribution.

Now we will make a special choice for $\rho(J)$ and see how it is
related to Wiener noise. Take $\rho(J)=\delta(J) +(1/2)
\delta^{\prime\prime}(J)$. Though it seems to be a very ridiculous
choice (for example, it can not be positive everywhere), but as we
shall see it lead to a welldefined noise for $x$.

Let's focus on the properties of $x$, such as its expectation
value and correlation functions. The equation (\ref{Xdefdel}) can
be solved to find $x(t)$:
\begin{equation}\label{XSol}
x(t)=\sum_i J_i \theta(t-t_i).
\end{equation}
It is clear from form of the distribution function we have chosen,
that the mean value of $x$ vanishes. Now consider the two point
correlation function $\langle x(t)x(t')\rangle$. Using the
solution (\ref{XSol}), we see
\begin{eqnarray}\label{2point1}
\langle x(t)x(t')\rangle = \left\langle \sum_{i,j} \left\langle
J_i J_j\right\rangle_J
\,\,\theta(t-t_i)\theta(t'-t_j)\right\rangle_t,
\end{eqnarray}
where the inner averaging is on different jump magnitudes and the
outer one is on distribution of time of jumps. As we have assumed
that magnitude of jumps are independent of one another, the inner
averaging vanishes unless if $i=j$. In this case, one should
compute the integral $\displaystyle{\int J^2\left[\delta(J) +
(1/2) \delta''(J)\right]}$ which is equal to one. So, the inner
averaging yields $\delta_{ij}$. Taking the time scale of the
poison distribution of time intervals of jumps to be unity, it is
easy to show the two point function (\ref{2point1}) turns out to
be
\begin{equation}\label{2point2}
\langle x(t)x(t')\rangle=\min(t,t'),
\end{equation}
which is the same as two point function of Wiener noise. It is
also easy to show that one can use Wick theorem to derive
$n$-point correlation functions. To have a more complete proof of
equivalence of the two noises, one can look at their Fokker-Plank
equation. Ito's formula for the general process (\ref{Xdefdel})
with $J$'s having distribution function $\rho(J)$, has the
following form:
\begin{eqnarray}\label{ITO1}
\frac{dF(x,t)}{dt}=\frac{\partial F(x,t)}{\partial
t}+\int\left[\left(F(x+J,t)-F(x,t)\right)\rho(J)\right]dJ
\end{eqnarray}
To derive this expression, first we should note that at any jump
$x(t)$ goes to $x(t)+J$ and also the fact that the the jump
magnitudes are independent of the poisson jump occurrence process,
that is
\begin{equation}\label{XIto}
\langle\left[F(x+J,t)-F(x,t)\right]dN\rangle=\langle\left[F(x+J,t)-F(x,t)\right]\rangle
dt
\end{equation}
where the averaging is over the probability distribution of jumps.
Also note that in equation (\ref{ITO1}), we have taken the time
scale of the poisson distribution to be unity.

Let's consider our specific choice of $\rho(J)$. Doing the
integration in equation (\ref{ITO1}) one can easily find the
related Fokker-Plank function
\begin{eqnarray}\label{Fokker}
\frac{\partial P(x,t)}{\partial t}+\frac 1 2\frac{\partial^2
P(x,t)}{\partial x^2}=0
\end{eqnarray}
which is the same as Fokker-Plank equation of Wiener noise.

One would criticize the equivalence of the two noises, telling
Wiener noise is a continuous one, but a noise having jumps is not.
However with the choice we have made, the process produces a
continuous noise, as one can examine that it satisfies Lindberg's
condition \cite{Gar}. The other point is that the probability
distributions we have chosen, are somehow odd, as they have
negative values at some points. Perhaps one can think of $\rho(J)$
as not a probability distribution, rather as a measure defined in
this way. This will also apply to other choices we'll mention in
below.

Before examining other choices for $\rho(J)$, we would like to
emphasize that, in SLE equation, one can use jumps with the
distribution mentioned above instead of Wiener noise and derive
the very same results, e. g. one can find the same Ito's formula
and the same null vectors, that is, all the results derived using
equation (\ref{SLEMain2}) could be re-derived using the same
equation, replacing Wiener noise with jumps which obey the above
distribution.

Now let's see what happens if we consider other distribution
functions such as third or higher derivatives of delta functions.
Take $\rho(J)=\delta(J) +(1/n!) \delta^{(n)}(J)$, where
$\delta^{(n)}(J)$ is the $n$'th derivative of delta function. The
solution to stochastic equation (\ref{Xdefdel}) is again given by
equation (\ref{XSol}), but now all the $m$-point functions
$\langle x(t_1)x(t_2)\ldots x(t_m)\rangle$ vanish for $m<n$ and
for $m=n$ one has
\begin{equation}\label{Npoint}
\left\langle x(t_1)x(t_2)\ldots
x(t_n)\right\rangle=\min(t_1,t_2,\ldots t_n).
\end{equation}

The Fokker-Plank of such evolution is just like equation
(\ref{Fokker}), with the second order derivation being replaced by
$n$'th order and the prefactor a half being replaced by
$\displaystyle{\frac{(-1)^n}{n!}}$. This allows us to extend the
ordinary SLE to a more general family of stochastic evolution. In
the next section we will see how this extension helps us to find
relations between SLE and CFT's with null vectors of third or
higher ranks.

\section{SLE and Higher Level Null Vectors}
In section 2, we saw that if $\xi_{t}$ be a Brownian motion then
we were able to define an infinite set of SLE zero modes, or
martingales, whose existence is a consequence of the existence of
a null vector at level two. Some CFT's do  not have null vectors
at level two, instead they have null vectors at higher levels. The
first question that may arise is that is it possible to find some
martingales which are related to higher null vectors so that SLE
could be related to such models. The ordinary SLE can not produce
such null vectors, because Brownian motion is only  able to
produce one and second order differentiations. But as we saw in
the previous section, replacing $\xi_{t}$ with jumps, which have
some specific distributions, then higher differentiations, and
hence the required operators to produce higher level null vectors,
appear.

Consider the following stochastic differential equation with
respect to a compensated poisson process:
\begin{equation}\label{SLEMod}
df_{t}=a(f_{t})dt+b(f_{t})dB_{t}+c(f_{t})J\,dN
\end{equation}
Taking $a(f)=2/f$ and $b(f)=\sqrt{k}$ and $c(f)=0$, one arrives at
the ordinary SLE$_k$ process. The last term is the one which
produces jumps. Note that the magnitude of the jump is given by
$c(f_t)\times J$.

Ito's formula for the process (\ref{SLEMod}) has the following
form:
\begin{eqnarray}\label{ITO}
dF(f_{t})&=&\left(a(f_{t})\partial
F(f_{t})+\frac{b(f_t)^2}{2}\partial^{2}F(f_{t})+\mathcal{L}\left[F(f_{t})\right]\right)dt+b(f_{t})\partial
F(f_{t})dB_{t}\\
\mathcal{L}\left[F(f)\right]&=&\int\left[\left(F(f+c(f)J)-F(f)\right)\rho(J)\right]dJ
\end{eqnarray}
The steps are the same as the ones in previous section, again the
time distribution is a poisson one with time scale equal to unity,
and the magnitude distributions is given by $\rho(J)$.

Examining  different distributions for $J$ and different
dependencies of $c$ on $f$, one is able to produce many different
$\mathcal{L}$'s. For example take $c(f_{t})=1$ and
$\rho(J)=\delta(J)$ then $\mathcal{L}F(f)=0$, that is we have not
added any jumps. But if we take
$\rho(J)=\delta(J)+\frac{\partial^{n}}{\partial J^{n}}\delta(J)$
keeping $c(f_{t})=1$, the resulting operator would be
\begin{equation}\label{diffn}
\mathcal{L}F(f)=(-1)^n\frac{\partial^{n}}{\partial f^{n}}F(f)
\end{equation}
This distribution helps us to produce the $L_{-1}^{n}$ operators
in Virasoro algebra. Another interesting choice, which is more
general, is $c(f_{t})=f_{t}^{m}$ and
$\rho(J)=\delta(J)+\frac{\partial^{n}}{\partial J^{n}}\delta(J)$,
which leads to the operators
\begin{equation}\label{mdiffn}
\mathcal{L}F(f)=(-1)^n\, f^m\frac{\partial^{n}}{\partial
f^{n}}F(f)
\end{equation}
which are the most general form of the operators needed to
construct higher level null vectors, as they can produce the
general form of operators in Virasoro algebra
$L_{n_{1}}L_{n_{2}}...L_{n_{k}}$.

Let's try to connect the modified Loewner's equation to CFT's
having null vector at level 3. Consider the following stochastic
evolution
\begin{equation}\label{StocLevel3}
df_{t}(z)=\frac{2}{f_{t}(z)^{2}}dt-\frac{2\sqrt{k}}{3f_{t}(z)^{\frac{1
}{2}}}dB+\alpha JdN
\end{equation}
Note that the maps are not defined in the upper half plane, they
are defined in only two third of the whole complex plane, say from
$\theta=0$ to $\theta=3\pi/2$ \cite{rasmussen}. Let
$\rho(J)=\delta(J)+\frac{\partial^{3}}{\partial J^{3}}\delta(J)$
and $\displaystyle{k=\frac{18}{h-1}}$,
$\displaystyle{\alpha=\frac{-2}{(h+2)(h-1)}}$. Writing Ito's
formula (equation(\ref{ITO})) and doing the same steps as in the
ordinary SLE case, one arrives at
\begin{equation}
\gamma_{f_t}^{-1} d\gamma_{f_t}=
c\,\left(l_{-3}-\frac{2}{h+1}l_{-1}l_{-2}+\frac{l_{-1}^3}{(h+1)(h+2)}\right)dt
+\frac{2\sqrt{k}}{3f_{t}(z)^{\frac{1 }{2}}}l_{-1}dB
\end{equation}
where $c= 2(h+1)/(1-h)$. Averaging over Brownian motion and going
to quantum level, will lead us to an operator which produces a
level three null vector
\begin{equation}\label{SLEMain9}
\left(L_{-3}-\frac{2}{h+1}L_{-1}L_{-2}+\frac{L_{-1}^3}{(h+1)(h+2)}\right)|\omega_{1,3}\rangle
=0
\end{equation}
where $|\omega_{1,3}\rangle$ is the highest weight vector with
weight equal to $h$ in a CFT which has a null vector at level 3.

One can do the same manipulations to relate the null vector at
level four to a modified SLE. In this case, one should take into
account the following stochastic equation:
\begin{equation}\label{SLEMain10}
df_{t}(z)=\frac{2}{f_{t}(z)^{3}}dt-\frac{\sqrt{k}}{2f_{t}(z)
}dB-\frac{p}{2f_{t}(z)}J_{1}dN_{1}+qJ_{2}dN_{2}
\end{equation}
Together with the following conditions, the  vector $G|\omega
_{1,4}\rangle$ is a martingale, where $G$ is the proper operator
derived from this evolution.
\begin{eqnarray}\label{Parameters}
\rho(J_{1})=\delta(J_{1})+\frac{\partial^{3}}{\partial
J_{1}^{3}}\delta(J_{1}),\hspace{26mm}
\rho(J_{2})=\delta(J_{2})+\frac{\partial^{4}}{\partial
J_{2}^{4}}\delta(J_{2})\hspace{7mm}\nonumber\\
 q=\frac{-9}{2(4h^2+9h +9)},\hspace{1cm}
p=\frac{-6(2h+3)}{4h^2+9h +9}\hspace{1cm} k=\frac{8(8h^2 +12 h +
9)}{4h^2+9h +9}
\end{eqnarray}
This means that we have a null vector at level four of the form
\begin{equation}\label{SLEMain11}
\left(L_{-4}-\frac{4h}{9}L_{-2}^{2}-\frac{4h+15}{6h+18}L_{-1}L_{-3}
+\frac{2h+3}{3h+9}L_{-1}^{2}L_{-2}-\frac{1}{4h+12}L_{-1}^{4}\right)\mid\omega_{2,2}\rangle
=0
\end{equation}
Note that at this case, the evolution is defined in the quarter of
complex plane.

Now it is clear that for any null vector at any level, we are able
to define a stochastic evolution, whose martingales could be found
using the properties of the null vector and the stochastic
evolution on Virasoro group.

Note that it is not the only possible way to produce desired
operators. We have had several assumptions which can be modified,
for example the assumption that the time distribution of jumps is
poisson distribution or the  magnitude of jumps are independent.
In general, one can define stochastic differential equations with
poisson point form jumps, with a specified form for the
distribution of jumps, which produce the  same differential
operators (see \cite{bass}, and references therein).

\end{document}